\documentclass[prl,twocolumn,showpacs,epsfig,preprintnumbers,amsmath,amssymb]{revtex4}

\usepackage{graphicx}% Include figure files
\usepackage{dcolumn}% Align table columns on decimal point
\usepackage{bm}% bold math

\def \be{\begin{equation}}
\def \ee{\end{equation}}
\def \bea{\begin{eqnarray}}
\def \eea{\end{eqnarray}}
\def \ba{\begin{array}}
\def \ea{\end{array}}

\begin{document}

%%%%%%%%%%%%%%%%%%%%%%%%%%%%%%%%%%%%%%%%%%%%%%%%%%%%%%%%%%%%%%%%%%%%%%%%%
%%%%%%%%%%%%%%%%%%%%%%%%%%%%%%%%%%%%%%%%%%%%%%%%%%%%%%%%%%%%%%%%%%%%%%%%%

\title{Correlation and response in a driven dissipative model}
\author{Dana Levanony and Dov Levine}
\affiliation{Department of Physics, Technion, Haifa 32000, Israel}
\date{\today}

\begin{abstract}
We consider a simple dissipative system with spatial structure in
contact with a heat bath.  The system always exhibits correlations
except in the cases of zero and maximal dissipation.  We explicitly calculate the
correlation function and the nonlocal response function
of the system and show that they have the same spatial dependence.
Finally, we examine heat transfer in the model, which agrees qualitatively with
simulations of vibrated granular gases.
\end{abstract}


 %Use showkeys class option if keyword display desired
\maketitle

%%%%%%%%%%%%%%%%%%%%%%%%%%%%%%%%%%%
%%%%%%%%%introduction%%%%%%%%%%%%%%
%%%%%%%%%%%%%%%%%%%%%%%%%%%%%%%%%%%

Driven dissipative systems (DDS) occur in many different contexts,
from collections of macroscopic particles to biological systems.
Such systems are intrinsically out of equilibrium, and need the
input of energy in order to remain functional.   One prototype DDS
is a granular gas\cite{granular_book}, a collection of inelastic
grains which dissipate energy through collisions; these have been
examined extensively, both experimentally\cite{gollub,menon},
numerically\cite{barrat,herbst,puglisi}, and
analytically\cite{kinetic_theory}. Because of the difficulty in
treating granular gasses analytically, stochastic mean-field models
have been studied.  One such is the Maxwell
model\cite{maxwell_model}, which assumes a velocity independent
collision rate and no spatial structure; this facilitates analytical
calculation \cite{naim} of quantities such as the  velocity
distribution function and its moments.

The introduction of spatial dependence complicates and enriches the
behavior, and may lead to correlations: for example, actual granular
gasses exhibit spatial clustering and velocity correlations because
of the dissipative collisions, as is seen in simulational studies  \cite{correlations}.
Williams and
MacKintosh\cite{mackintosh} have shown numerically that correlations
exist in a 1D driven dissipative gas provided the restitution
coefficient is different from one, and Soto, {\it et al}\cite{soto}
have used the BBGKY hierarchy to study the appearance of velocity
correlations in inelastic hard-sphere systems. Baldassarri, {\it et al}\cite{bald}, and Ben-Naim
and Krapivsky\cite{bennaim} consider a lattice variant of the
Maxwell model which they solve in the freely cooling case (no
driving); the latter authors calculate the spatially dependent velocity correlations
which exhibit gaussian decay with distance.

In this Rapid Communication we study a model of a driven system with
spatial structure: the constituent ``particles" are constrained to
lie on a 1D lattice with nearest-neighbor coupling. Our main goal
will be to understand the connection between the system's
dissipative nature and spatial correlations.  The  system is coupled
to a heat reservoir at temperature $T$, and the model is chosen so
that it has a well-defined equilibrium limit for certain values of
the system parameters.  The mean-field version of the model, which
has no spatial structure, can be solved exactly\cite{Dov&Yair}, in
the sense that all the moments of the energy distribution may be
calculated.   For the model of this Rapid Communication, we
calculate the two-point correlations of the system analytically, and
demonstrate that non-zero correlations always exist except for the
cases (a) in which there is no dissipation (in agreement with the
results of \cite{peng,mackintosh}) or (b) when the dissipation is
maximal. For dissipative systems driven by thermal contact with a
heat bath, it is not only the bath temperature which determines the
steady-state of the system; the nature of the coupling to the bath
is relevant as well (in contrast to non-dissipative systems, for
which this last plays no role in the determination of the
equilibrium state).
 With this in mind, we calculate the response of the
system to a change in one of these defining parameters, and compare
the spatial dependence of the nonlocal response function to that of
the correlation function. Finally, motivated by the question of
energy flow in vibrated granular materials\cite{barrat,herbst,
heat_simu} we study how the internal dissipation affects the nature
of heat transfer, and calculate the energy along a chain whose ends
are in contact with a heat bath. We note that the model studied in
reference \cite{linear_solution} in this context is the same as our
model in the absence of dissipation.

%%%%%%%%%%%%%%%%%%%%%%%%%%%%%%%%%%%%%%%%%%%%%%%%%%%%%%%%%%%%%%%%%%%%%%%%%
%The model and solution of 1dimension
%%%%%%%%%%%%%%%%%%%%%%%%%%%%%%%%%%%%%%%%%%%%%%%%%%%%%%%%%%%%%%%%%%%%%%%%%
%\section{The one-dimensional lattice}
The system we study is a generalization of a model introduced in
\cite{Dov&Yair}, which can be regarded as the mean-field version of
the case of this Rapid Communication. We consider N particles
localized on sites $n$ of a 1D lattice, each characterized by its
energy $E_{n}$. The entire chain is coupled to an external
Boltzmann-distributed bath with temperature T. At each time step an
interaction occurs in the system, either between system particles or
with the external bath. Specifically, a particle is chosen randomly
and its interaction follows the stochastic rule :
\bea E_n(t+dt)= \left\{
\begin{array}{cc}
  \underline{value}: & \underline{probability:} \\
  E_n(t) & 1- \Gamma dt \\
  z \alpha (E_n(t)+E_{n+1}(t))  & \frac{1}{2} (1-f) \Gamma dt \\
  z \alpha (E_n(t)+E_{n-1}(t))  & \frac{1}{2} (1-f) \Gamma dt \\
  z (E_n(t)+E_{B})  & f \Gamma dt
\end{array} \right.\label{eq:dyn_rule}
\eea
Here $ \Gamma$ is the overall rate of interaction of a particle;
it sets the time scale and is irrelevant to the steady state.
$f$, the strength of the coupling to the bath, is a constant which determines the probability of a particle
interacting with the bath, $\alpha \in [0,1]$ is a parameter
characterizing the dissipation in an interaction (in analogy to a
restitution coefficient), $z$ is a stochastic variable uniformly
distributed between 0 and 1; and $E_{B}$ is the energy of a
particle chosen randomly from the bath.  In what follows, we
shall only be interested in the steady states of the system.

Our main results are: (i) correlations appear in this system for all
$0 < f < 1$ provided $\alpha <1$; when f = 0,1 or $ \alpha =1$ there
are no correlations; (ii) the spatial decay of the correlation
function is calculated and found to be exponential in the limit of $
N \rightarrow \infty $; (iii) the nonlocal response function (to a
localized change in $f$ or $T$) is proportional to the correlation
function.  This last point is reminiscent of the
fluctuation-dissipation theorem of equilibrium statistical
mechanics, but in our case there is no {\it a priori} reason to
expect that the two functions will exhibit the same spatial
dependence.

%%%%%%%%%%%%%%%%%%%%%%%%%%%%%%%%%%%%%%%%%%%%%%%%%%
%%%%%%%%%%%%%%%%%calculating the correlations%%%%%
%%%%%%%%%%%%%%%%%%%%%%%%%%%%%%%%%%%%%%%%%%%%%%%%%%

\section{Correlation and Response}
Using the dynamics presented in (\ref{eq:dyn_rule}) we can write
equations describing the time evolution of the moments of the energy distribution and the
correlation functions in the system.  We will be interested in the
steady-state values of the moments, so we set the time derivatives
to zero. It is simple to show that the average energy is given by
$\overline{E_{n}}\equiv\overline{E}= \frac{Tf}{2-f-2\alpha (1-f)}$,
with $T$ being the bath temperature in units where the Boltzmann
constant is unity. In order to compute the correlation function
$C(n,n+k)\equiv\overline{E_{n}E_{n+k}}-\overline{E}^{2}$, we need to
calculate the second moments; this leads to a set of coupled
equations:
$$
 \overline{E^{2}}= \frac{2 \alpha
^{2}(1-f) \overline{E_{i}E_{i+1}} +2Tf\overline{E}+2fT^{2}}{3-2 \alpha ^{2} (1-f)-f}
 \label{eq:sec_moment}
$$

$$
\overline{E_{n}E_{n+1}} = \frac{6fT\overline{E}+(1-f) \alpha
 (\alpha \overline{E^{2}}+3\overline{E_{n}E_{n+2}})}{9-3f-\alpha(1-f)(3+\alpha)}
  \label{eq:fir_corr}
$$

$$
{\overline{E_{n}E_{n+k}}}= \frac{\frac{1}{2} \alpha (1-f)
(\overline{E_{n}E_{n+k+1}}+\overline{E_{n}E_{n+k-1}})+fT\overline{E}}{2-
\alpha (1-f)-f}. \label{eq:corr}
$$
In the last equation, $k \geq 2$.  We note that if $f=0,1$ or $
\alpha =1 $, then $C(n,n+k)=0$: there are no correlations in these
cases. This is consistent with the two-dimensional granular gas
simulation of \cite{peng} where velocity correlations disappear as
the restitution coefficient goes to 1, and reminiscent of similar
behavior of the density correlations\cite{mackintosh}. The case $
\alpha =0$ is unique in the sense that we obtain the mean field
result for the distribution function and all correlations disappear
except $C(n,n+1)$.

% $<E_{i}E_{i+1}>-<E>^{2}=\frac{<E>^{2}(1-f)}{3-f}$ .

For general $f$ and $\alpha$, the above coupled equations can be
written as a matrix equation of the form $\cal
A\overrightarrow{W}=\overrightarrow{V}$, where $\overrightarrow{W}$
is a vector whose $k^{{th}}$ place is $\overline{E_{n}E_{n+k}}$, and
$\overrightarrow{V}$ is a vector of constants.
The $N \times N$ matrix ${\cal A}$ may be decomposed as
$\cal A = \cal T - \cal B$, where $\cal T$ is a tridiagonal matrix with constant
diagonals (which can be inverted with the help of \cite{inverting_matrix}), and $\cal B$
is an $N \times N$ matrix which is zero everywhere but the upper left $3 \times 3$ block
(which is denoted by the $3 \times 3$ matrix $\cal G$). ${\cal A}^{{-1}}$ may be computed
from the relation\cite{dana&dov2}
\begin{equation}
{\cal A} ^{-1} = {\cal T} ^{-1} + {\cal T}^{-1} {\cal B} {\cal Q} {\cal T}^{-1}
\end{equation}
where $\cal Q$ is an $N \times N$ matrix of zeros except for the
upper left $3 \times 3$ block which is the matrix $({\cal I} - {\cal
G})^{-1}$, $\cal I$ being the $3 \times 3$ unit matrix.
In the limit $N\rightarrow \infty$ this yields the result that the
correlations decay as \bea C(n,n+k)=D(\alpha ,f) e^{
-\frac{k}{\lambda}} \label{eq:correlation} \eea where $D(\alpha ,f)$
is a continuous function,  and  $\lambda$ is the correlation length,
given by $\lambda ^{-1} =arccosh \frac{2- \alpha
(1-f)-f}{\alpha(1-f)}$.

%%%%%%%%%%%%%%%%%%%%%%%%%%%%%%%%%%%%%%%%%%%%%%%%%%%%
%%%%%%%correlation length%%%%%%%%%%%%%%%%%%%%%%%%%%%
%%%%%%%%%%%%%%%%%%%%%%%%%%%%%%%%%%%%%%%%%%%%%%%%%%%%

%\section{correlation length and the near equilibrium regime}

\begin{figure}[htb]
\includegraphics[width=8cm]{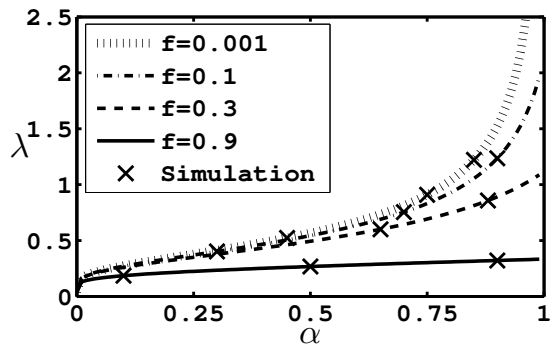}
\caption{\label{fig:length} The correlation length as a function of
$\alpha$ for different values of $f$.}
\end{figure}

We note that the correlation length diverges as $ \alpha \rightarrow
1 $ and $f \rightarrow 0 $, although we know that for $ \alpha =1 $
there are no correlations; this is because $D(\alpha , f)
\rightarrow 0$ for these values. This means that around $\alpha=1$
and $f=0$ the correlations are the longest-ranged but the weakest.
The behavior of the correlation length as a function of $f$ and
$\alpha$ is shown in Figure \ref{fig:length}, in which results from simulating the
model are presented for comparison.

%%%%%%%%%%%%%%%%%%%%%%%%%%%%
%%%figure correlation length
%%%%%%%%%%%%%%%%%%%%%%%%%%%%

%%%%%%%%%%%%%%%%%%%%%%%%%%%%%%%%
%%%%%%%%distribution function%%%
%%%%%%%%%%%%%%%%%%%%%%%%%%%%%%%%

%\section{distribution function}
Because of the correlations, the single-particle energy distribution function $P(E)$
cannot be calculated analytically for general $ \alpha$. For $\alpha
=1$, detailed-balance holds, and an H-theorem may be proved,
and the system comes to thermal equilibrium at the bath
temperature $T $. When $ \alpha =0$, the generating function
$g(\omega) $ (defined by $g(\omega) \equiv \langle e^{-\omega E}
\rangle = \int_0^{\infty} e^{-\omega E} P(E) dE $) is the same as
that of the mean field model \cite{Dov&Yair}: $ g(\omega)= (\omega
T+1) { _2F_1 }(1,2,2-f,- \omega T)  $ where $_2F_1$ is a
hypergeometric function.

\begin{figure}[htb]
\includegraphics[width=8cm]{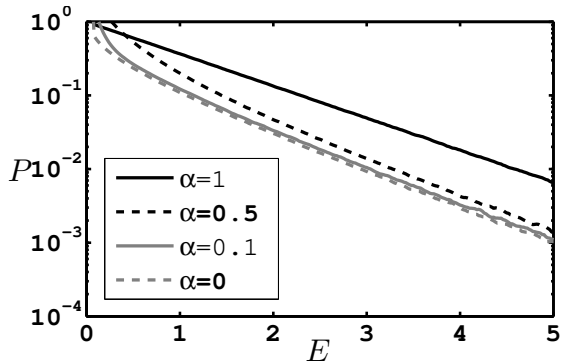}
\caption{\label{fig:distribution} The distribution function on a
logarithmic scale as found from Monte Carlo simulation with  $f=0.5$
and $T = 1$.}
\end{figure}

For  $f=0$ (and $\alpha \neq 1)$  the system energy decays to zero
and therefore $P(E)= \delta (0)$ is the trivial steady-state.  In
Figure \ref{fig:distribution}, we plot the energy distribution
function $P(E)$ for different values of $\alpha$.  We note that the distributions in Figure
\ref{fig:distribution} are reminiscent of those found for the
mean-field case\cite{Dov&Yair}.
%%%%%%%%%%%%%%%%%%%%%%%%%%%%%%%%%%%%%
%%%%%%figure distribution function%%%
%%%%%%%%%%%%%%%%%%%%%%%%%%%%%%%%%%%%%

%%%%%%%%%%%%%%%%%%%%%%%%%%%%%%%%%%%%%%%%%%%%%%%%
%%%%%%%%%%%%%%%response function%%%%%%%%%%%%%%%%
%%%%%%%%%%%%%%%%%%%%%%%%%%%%%%%%%%%%%%%%%%%%%%%%

%\section{Response function}
For a dissipative system coupled to a bath, both the coupling
strength $f$ and the bath temperature $T$ determine the behavior of
the system. We will now consider the response of the system to a
local change in the coupling to the bath. We stress that this is not
a measurement which can be performed on a system in equilibrium, for
which $f$ plays no role in the steady state, and it is thus
intrinsically a non-equilibrium measurement. Nevertheless, we shall
see that, reminiscent of fluctuation-dissipation relations, the spatial
response to such a perturbation is proportional to the spatial
correlation function.

We imagine that each site in the system is coupled to a
Boltzmann-distributed bath at temperature $T$ with a coupling
strength $f$.  We seek to calculate the change of energy at site
$n+k$ due to small change in $f$ (say, from $f$ to $f_{0}$) at
location $n$. Having changed the coupling at site $n$, the system no
longer has spatial invariance, so instead of writing the
time-evolution equations for a single particle, we must consider the
dynamics of the entire chain. The structure of the resulting
equation\cite{dana&dov2} is again  of the matrix form ${\cal
A}\overrightarrow{E}=\overrightarrow{V}$, where $\overrightarrow{E}$
is a vector of the energies $\{E_{j}\}$, and $\overrightarrow{V}$ is
a vector of constants related to the boundary conditions.  ${\cal
A}$ is a matrix which depends on $\alpha, f$ and $f_0$ which can be
inverted to yield ${\overline E_{n+k}}$, from which we obtain that,
in the $N\rightarrow\infty$ limit,
\bea
(\frac{d{\overline E_{n+k}}}{df_{0}})_{f_{0}=f}=B(\alpha,f)
e^{-k/\lambda}
 \eea
where $B(\alpha,f)$ is a continuous function, and where $\lambda$ has the same
value as in
Equation \ref{eq:correlation}. This describes the nonlocal response of the system
at a site a distance $k$ away from the point of a local change in coupling to the
bath.  We note that this
intrinsically non-equilibrium response has the same spatial decay as
correlation function.

We may similarly ask what the response at a site $n+k$ is to a local
change in temperature $T$ at site $n$.  As before, we are able to
invert the resulting matrix equation, which, in the limit
$N\rightarrow\infty$
% yields an integral, which may be evaluated
\cite{dana&dov2} gives:
\bea \frac{d{\overline E_{n+k}}}{dT_{n}}=
\frac{f}{\alpha (1-f) sinh(\lambda^{-1})} e^{-k/\lambda}
\label{eq:response_T} \eea

In an equilibrium system, the response of an observable to a small
change in its conjugate field is proportional to the correlations of
this variable, with the system's temperature being the constant of
proportionality.  Although our response measurements do not take
this form, it is interesting to note that in both cases, the
response of the energy ${\overline E_{n+k}}$ to a local change in
the bath interaction at site $n$ is proportional to the correlation
function $C_{n,n+k}$.  Moreover, for given bath temperature, the
ratios of the response to the correlation (for both types of
response) decrease monotonically with $\alpha$, and diverge as
$\alpha \rightarrow 0$.

%%%%%%%%%%%%%%%%%%%%%%%%%%%%%%%%%%%%%%%%%
%%%%%%%%%%%heat transfer%%%%%%%%%%%%%%%%%

%%%%%%%%%%%%%%%%%%%%%%%%%%%%%%%%%%%%%%%%%

 \section{heat transfer}
We have thus far considered a system  all of whose sites are in
contact with the same external bath.  As an application of the ideas
of the previous section, we now turn our attention to the case where
the ends of the lattice are coupled to heat baths at temperatures
$T_{-}$ and $T_{+}$, respectively. For the case of $\alpha = 1$,
we expect that the temperature
profile ({\it i.e.} the average energy at a site) will be a linear
interpolation between the two wall temperatures; this is a
manifestation of the Fourier law\cite{linear_solution,heat_equi,
heat_simu, fourier}.

\begin{figure}[htb]
\includegraphics[width=8cm]{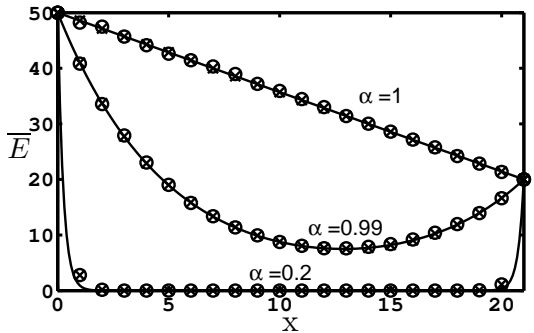}
\caption{\label{fig:result1} Average energy as a function of position for a 20 particle
lattice for different values of $\alpha$.  The symbols $\circ$ are from a simulation of the
model, the values from $\cdot$ from Equation \ref{eq:solution}, and the continuous curve
is the solution of the damped diffusion equation obtained in the continuum limit.}
\end{figure}

The case where $\alpha \neq 1$ may be thought of as an idealization
of energy transport through an inhomogeneous driven granular gas,
where we might imagine causing walls to vibrate and ask how this is
manifested throughout the system. For example, Herbst, {\it et al}
\cite{herbst} and Barrat, {\it et al} \cite{barrat} treat a
collection of inelastic disks in 2D held between vibrating walls,
and find a characteristic decay of the granular temperature with
distance from the wall.  We modify our system so that of the
particles on the lattice, only those at the end ($k = 0, N+1$) are
coupled to baths, at temperatures  $T_-$ and $T_+$, respectively.
For all other particles,  the dynamics are as in Equation
\ref{eq:dyn_rule}, with $f = 0$. In similar fashion to the previous
calculation, the relation for the average energy at site $k$ may be
expressed as a matrix equation, which may be inverted to give
 \bea
{\overline  E_{k}}=\frac{T_{+}sinh[(k-N-1)\eta] +T_{-}
sinh[k\eta]}{sinh[(N+1)\eta]} \label{eq:solution}
 \eea
with $\eta=arccosh\frac{2-\alpha}{\alpha}$. This result is plotted
in Figure \ref{fig:result1}. We note that for $\alpha \rightarrow
1$, the profile is linear, as expected\cite{linear_solution}.  For
all other $\alpha$, sufficiently far from a wall the decay is
exponential with decay length $\eta$.  It is interesting to note the
similarity of these results to those of the numerical study of
driven granular gasses in \cite{barrat,herbst}.

This result may be understood in the context of the Fourier
law\cite{fourier} which relates the rate of heat flow,
$\frac{dQ}{dt}$ to the local temperature gradient:
$\frac{dQ}{dt}=-\kappa \frac{d^2T}{dx^2}$.  In our case, the energy
$E_{i}$ plays the role of both $Q$ and $T$, and its time evolution
is governed by:
 \bea\ba{c} \frac{dE_i(t)}{\Gamma dt}=\frac{2E_i(\alpha
-1)}{N+1}+\underset{J_+}{\underbrace{\frac{\alpha}{2(N+1)}(E_{i+1}-E_i)}}\\
-\underset{J_-}{\underbrace{\frac{\alpha}{2(N+1)}(E_{i}-E_{i-1})}}
\ea\eea
The terms on the right hand side of this equation can be interpreted as
a sink term, and the heat flow out of, and into site $i$ respectively.
The net average energy flux into $i$ is $J_{-}= -\kappa
(\overline{E_{i}}-\overline{E_{i-1}})$ with heat conductivity
coefficient $\kappa= \frac{\alpha}{2(N+1)}$.
We note that in the continuum limit we obtain a damped diffusion equation for the steady-state:
$\alpha \nabla
^{2}E+4(1-\alpha)E=0$
with boundary condition $E(x=0)=\alpha
T_{+}$, $E(x=N+1)=\alpha T_{-}$.
\\

In this Rapid Communication, we have described exact calculations
for the spatial dependence of correlations and response for a model
DDS.  It is intriguing that, despite the fact that the
fluctuation-dissipation theorem is not applicable to this system,
the nonlocal response to a change in temperature or bath coupling
has the same spatial dependence as the correlation function.  Of
course, it remains to be seen what the nature of the temporal
behavior of these functions is, and it would be surprising if they
were to exhibit the same frequency dependence.

%%%%%%%%%%%%%%%%%%%%%%%%%%%%%%%%%%%%%%%%%%%%%%%%%%%%%%%%%%%%%%%%%%%%%%%%%
%Acknowledgments:
%%%%%%%%%%%%%%%%%%%%%%%%%%%%%%%%%%%%%%%%%%%%%%%%%%%%%%%%%%%%%%%%%%%%%%%%%

We would like to thank Joseph Avron, Guy Bunin, Jean-Pierre Eckmann,
Joshua Feinberg,  Michael Fisher, Oded Kenneth, Yair Srebro, and
Annette Zippelius for interesting and enlightening discussions. D.
Levine  gratefully acknowledges funding from Israel Science
Foundation grants  88/02 and 660/05, and the Technion Fund for the
Promotion of Research, and thanks the New York University Physics
department for its hospitality.
%%%%%%%%%%%%%%%%%%%%%%%%%
%%%%%%References%%%%%%%%%
%%%%%%%%%%%%%%%%%%%%%%%%%

\end{document}